\documentstyle[prl,aps,twocolumn]{revtex}

\def\be{\begin{equation}}
\def\ee{\end{equation}}
\def\beqa{\begin{eqnarray}}
\def\eeqa{\end{eqnarray}}

\begin{document}
\title{ Comment on  ``Phase and Phase Diffusion of a Split Bose-Einstein 
Condensate"
}

\author{
A. J.~Leggett${}^1$$^*$ and F. Sols${}^2$ 
}
\address{
${}^1$ Laboratoire de Physique Statistique de l'Ecole Normale Superieure,
associ\`e au CNRS et aux Universit\`es Paris 6et 7 \\
24 rue Lhomond, F-75231 Paris Cedex, France \\
${}^2$ Departamento de F\'{\i}sica Te\'orica
de la Materia Condensada,
Universidad Aut\'onoma de Madrid, E-28049 Madrid, Spain 
}
\maketitle



Recently Javanainen and Wilkens \cite{java97} (hereafter JW) analysed
an experiment in which an interacting Bose condensate, after being allowed to
form in a single potential well, is ``cut" by splitting the well
adiabatically with a
very high potential barrier. They found that following the cut the rate $R$
at which
the two halves of the condensate lose the ``memory" of their relative phase is
$\sqrt{N}\xi$, where $\xi \equiv d^2E/dk^2$, $k$ being
the number of particles transferred between the two halves of the split
trap ($\hbar=1$). We show here that the problem posed by JW reduces to one
already studied in the literature \cite{ande86,legg91,sols94},
and that the true value of $R$ is much smaller.

We interpret JW's ``adiabatic condition" as the statement    
that the quantity 
$\lambda \equiv (dV/dt)/ \hbar \omega_0^2$ is $\ll 1$,
where $V$ is the barrier height  and $\omega_0$ the 
small-oscillation frequency in the unsplit well, while
simultaneously $r \equiv
\lambda \omega_0 \gg R$ where $R$ is the phase diffusion rate which we will
eventually calculate. We shall assume the condition  (experimentally
realistic for the BEC alkali gases and implicit in
JW's Eq. (11))  $Q \equiv (15Na/l)^{1/5} \gg 1$ (with
$a$ and $l$ defined as in JW).

  Let us consider a particular value of the barrier height $V_0$
large enough that the contact between the
two halves of the well is by Josephson tunnelling. We define the Josephson
coupling energy $E_J$ and the Josephson plasma frequency $\omega_p
\equiv (E_J \xi)^{1/2}$ in the
standard way; we remark that, for the BEC alkali gases, the conditions
$\omega_0 \gg 
\omega_p \gg R$ and $E_J \gg \xi$ 
can be simultaneously satisfied over a wide range of barrier heights
\cite{zapa97}.

We now first consider a rather artificial two-stage
process. At stage I we raise the barrier from zero to $V_0$, at a rate 
$r$ 
such that $R \ll r \ll \omega_p$.
Because of the second inequality, this process is indeed strictly
``adiabatic" and the final state of the system
is with negligible error the groundstate of the (interacting) Hamiltonian
for the final 
$V_0$. This in turn can be well approximated 
\cite{zapa97} by the 
standard Josephson ``pendulum" Hamiltonian (see e.g. Eq. (1) of Ref.
\cite{sols94}). The crucial point to note is that the
fluctuations, in the groundstate  \cite{comm1}  of this Hamiltonian, of the
relative particle 
number $k$ are by no means of order $\sqrt{N}$ but rather much smaller, of
order
$(E_J/ \xi)^{1/4}$.
The phase fluctuations, while 
finite, are correspondingly of order $(\xi/E_J)^{1/4} \ll 1$. Note that the
argument of this paragraph
does not depend on the (incorrect) assumption that Eq. (1) of Ref.
\cite{sols94} describes
the unsplit well. 

At stage II we raise the barrier height from $V_0$ to
``infinity" at a rate $r'$ such that $\omega_p \ll r' \ll  \omega_0$,
and then leave the system     
alone for a time $t$.. 
The problem of phase diffusion at this stage is precisely that considered
in Refs. \cite{ande86,legg91,sols94}; 
from Eq. (19) of Ref. \cite{legg91} we find
that the dephasing
rate $R$ is approximately
\be
R=(E_J \xi^3)^{1/4}.
\ee

We now turn to the more ``natural" case of a single smooth variation of the 
barrier height with time. It is clear that we should expect a formula of the
general form (1) still to apply, provided that $E_J$ is taken to be of the
order
of the Josephson energy at the point where the adiabatic condition is first
violated (i.e. when $|d\omega_p/dt| \sim \omega_p^2$), which (if  $\lambda
\ll 1$)
can be shown to lie well within the tunnelling regime.
Thus, the true dephasing rate in the experiment analysed by JW 
is always smaller than the result they give by a factor of order 
$N^{-1/2} (E_J/\xi)^{1/4}$. 
Using the fact \cite{zapa97} that $E_J \propto \exp(-S_0)$ 
and that the WKB exponent $S_0$ is a slowly varying      
factor times $V_0/\omega_0$, it is straightforward to show that up to
factors of order 
$| \ln \lambda|$ this ratio is of order $\sqrt{\lambda}/Q$.
We conclude that, if  $Q \gg 1$, JW's result cannot describe correctly an
adiabatic
($\lambda \ll 1$) separation process.
 
This work has been supported by DGICyT (PB93-1248) and by NSF
(DMR 96-14133). AJL would like to thank S. Balibar and his
colleagues 
for their warm hospitality at the ENS, Paris, where this work was performed.


\begin{thebibliography}{1}
\bibitem{comm1} 
Which is of course a superposition of Hartree-Fock states corresponding to
(slightly) different
values of the relative phase, see Eqs. (15) and (18) of Ref.  \cite{legg91}
and the intervening
discussion.
\end{thebibliography}
\end{document}